\documentclass[prl,aps,showpacs,twocolumn,preprintnumbers,nofootinbib,amsmath,amssymb,superscriptaddress,longbibliography]{revtex4-2}

\usepackage{systeme}
\usepackage{mathtools}
\usepackage[inline]{enumitem}
\usepackage{graphicx}

\usepackage[T1]{fontenc}
\usepackage[english]{babel}
\usepackage{xcolor}
\usepackage{braket}
\usepackage{nicefrac}
\usepackage{bm}
\usepackage{bbm}
\usepackage{braket}
\usepackage{bbold}    
\usepackage{slashed}
\usepackage[normalem]{ulem}
\usepackage{array}
\newcolumntype{C}{>{$}c<{$}}

\newcommand{\beqn}{\begin{eqnarray}}
\newcommand{\eeqn}{\end{eqnarray}}
\newcommand{\beqs}{\begin{subequations}}
\newcommand{\eeqs}{\end{subequations}\\[-2mm]\noindent}
\newcommand{\eq}[1]{(\ref{#1})}

\newcommand{\bs}{\boldsymbol}

\allowdisplaybreaks

\definecolor{brickred}{rgb}{0.8, 0.25, 0.33}
\definecolor{macouleur}{RGB}{105,150,150}

\usepackage{color}
\definecolor{purple}{rgb}{0.8,0,0.6}

\def\bbbone{{\mathchoice {\rm 1\mskip-4mu l} {\rm 1\mskip-4mu l} {\rm 1\mskip-4.5mu l} {\rm 1\mskip-5mu l}}}

\begin{document}

\bibliographystyle{apsrev4-1}

\title{Instantons in rotating finite-temperature Yang-Mills gas}

\author{M. N. Chernodub}
\affiliation{Institut Denis Poisson UMR 7013, Universit\'e de Tours, 37200 Tours, France}

\begin{abstract}
We find an instanton (caloron) solution in the finite-temperature SU(2) gluon gas subjected to (imaginary, in Euclidean spacetime) rotation. We demonstrate that the rotation decreases the temperature of the caloron and leads to the delocalization of its topological charge over fractionally charged constituents. Furthermore, we show that in the high-temperature limit, the rapidly-rotating caloron becomes a ``circulon'': a self-dual monopole (dyon) possessing a spatial toroidal core. 
\end{abstract}

\date{\today}

\maketitle


The recent observation of highly-vortical quark-gluon plasma~\cite{STAR:2017ckg} has attracted attention to the properties of relativistically rotating quark-gluon matter. The vorticity of the quark-gluon plasma can be probed experimentally via spin polarization of the hadronized matter~\cite{Becattini:2020ngo, Huang:2020dtn}. Theoretical analysis indicates that rotation can also affect the thermodynamic properties and modify the phase diagram of hot QCD~\cite{Chen:2015hfc, Jiang:2016wvv, Chernodub:2016kxh, Chernodub:2017ref, Sadooghi:2021upd, Wang:2018sur, Chen:2020ath, Zhang:2020hha, Braguta:2020biu, Chernodub:2020qah, Braguta:2021jgn, Fujimoto:2021xix,Chen:2022smf}. Although most of the theoretical studies were carried out in the much simplifying approximation of rigid rotation, a global picture of the phase diagram of rigidly-rotating quark-gluon plasma is still missing. 

In the absence of rotation, the finite-temperature QCD transition appears to unify the chiral and deconfining crossovers that occur close to each other~\cite{Bazavov:2011nk}. However, the theoretical results on chiral and deconfining transitions in rotating plasma contain certain controversies. 

All existing theoretical studies agree that vorticity reduces the critical temperature of the chiral phase transition~\cite{Chen:2015hfc, Jiang:2016wvv, Chernodub:2016kxh, Chernodub:2017ref, Sadooghi:2021upd, Wang:2018sur, Chen:2020ath, Zhang:2020hha}. The physical mechanism of this phenomenon takes its roots in the Barnett effect~\cite{Barnett:1915} discovered in 1915: the rotation aligns the spins of quarks and anti-quarks along the rotation axis and suppresses the pairing of fermions in the scalar channel, thus inhibiting the quark condensate~\cite{Jiang:2016wvv}. Thus, the faster rotation, the lower the chiral critical temperature.

However, the first-principle numerical simulations of purely gluonic matter show a different picture implying that the critical temperature of the deconfining transition should increase with the angular velocity~\cite{Braguta:2020biu, Braguta:2021jgn}: the faster rotation, the higher the deconfining critical temperature. At the same time, other independent approaches to the deconfinement problem in rotating quark-gluon plasma, a holographic technique of Ref.~\cite{Chen:2020ath} and the hadron-resonance gas used in Ref.~\cite{Fujimoto:2021xix}, indicate, in a contradictory manner as well, that the deconfining critical temperature should decrease with the angular velocity. Moreover, the Tolman-Ehrenfest argument, re-derived in a confining model in Ref.~\cite{Chernodub:2020qah}, suggests that rotation leads to yet another outcome: a qualitative change of the QCD phase diagram featuring a new mixed confining-deconfining phase with both phases present at the same time in a broad domain of the parameters of the model. Thus, the connection between the chiral and confining properties and intrinsic coherent explanation of the confining mechanism are missing.

Let us now look briefly at the problems of chiral symmetry breaking and confinement from the point of view of the topology of the Yang-Mills gauge group\footnote{Here, we do not consider alternative approaches based on partial gauge fixing of non-Abelian symmetry and subsequent use of topology of the residual Abelian symmetry group to uncover the topological content of the gluonic configurations~\cite{Mandelstam:1974pi,tHooft:1981bkw, DelDebbio:1996lih}.}. The natural topological objects of zero-temperature Euclidean Yang-Mills theory are instantons representing self-dual solutions to the classical Yang-Mills equations of motion~\cite{Belavin:1975fg}. The Euclidean instanton configuration can be associated with a process in a Minkowski spacetime.

The instanton is a solution to the classical equations of motion, $\partial_\mu F_{\mu\nu} + [A_\mu,F_{\mu\nu}] = 0$ for the Yang-Mills field $A_\mu = g t^a A^a_\mu$, where $t^a$ are the generators of the $SU(N_c)$ gauge group with $a=1,\dots N_c^2 -1$. The self-dual instanton solutions are determined by the relation ${\tilde F}_{\mu\nu} \equiv \epsilon_{\mu\nu\alpha\beta} {\tilde F}_{\alpha\beta} = \pm F_{\mu\nu}$ imposed on the field strength tensor $F_{\mu\nu} = \partial_\mu A_\nu - \partial_\nu A_\mu + [A_\mu,A_\nu]$. A convenient form of a general instanton solution is
\beqn
A^a_\mu = - {\bar \eta}^a_{\mu\nu} \partial_\nu \ln \phi\,,
\label{eq:A:a}
\eeqn
where ${\bar \eta}^a_{\mu\nu}$ is the 't~Hooft symbol~\cite{tHooft:1976snw} and $\phi$ is the scalar potential which satisfies the equation $\Box \phi = 0$. 

The topology of the Euclidean gluon configuration is characterized by the integer-valued Pontryagin index (topological charge),
\beqn
Q = \frac{1}{16 \pi^2} \int d^4 x \, {\mathrm{Tr}}\, \left( {\tilde F}_{\mu\nu} F_{\mu\nu} \right) 
= \pm \frac{1}{8 \pi^2} S\,,
\label{eq:Q}
\eeqn
where the Euclidean action is:
\beqn
S = \frac{1}{2 g^2} \int d^4 x \, \int d^4 x \, s(x),\qquad  s(x) = - \Box \Box \log \phi. \quad
\label{eq:S}
\eeqn

An $N$-instanton configuration is given by the potential:
\beqn
\phi = 1 + \sum_{k=0}^N \frac{\lambda_k^2}{(x - x_k)^2}\,,
\label{eq:phi:sum}
\eeqn
where $k^{\mathrm{th}}$ instanton is characterized by its size $\lambda_k$ and position $x_k$.

The role of instantons in QCD is impossible to overstate~\cite{Schafer:1996wv}. Instantons are believed to be responsible for the chiral symmetry breaking since they host exact fermionic zero modes. An instanton gas or liquid lifts the degeneracy of zero modes producing a finite density of near-zero-mode states that, in turn, lead straightforwardly to the breaking of the chiral symmetry~\cite{Diakonov:1985eg}. As temperature increases, the instantons form instantonic molecules (see, for example, Ref.~\cite{Ilgenfritz:1994nt}), which possess a gapped fermionic spectrum. Thus, the instanton picture of the QCD vacuum correctly implies the chiral symmetry breaking at low temperatures and reproduce the chiral symmetry restoration at high temperature.

At finite temperature $T$, the Euclidean time direction $\tau \equiv - i t$ is compactified into a circle $S^1$ of the length $\beta = 1/T$ with periodic boundary conditions imposed on the gluon fields, 
\beqn
A_\mu({\bs x},\tau) = A_\mu({\bs x},\tau + \beta)\,.
\label{eq:periodic:A}
\eeqn
At finite temperature, the instanton solutions become periodic instantons, so-called Harrington-Shepard (HS) calorons~\cite{Harrington:1976dj,Harrington:1978ve}. The periodicity of the gluon field~\eq{eq:periodic:A} translates into the periodicity of the scalar potential~\eq{eq:A:a}:
\beqn
\phi({\bs x},\tau + \beta) = \phi({\bs x},\tau)\,.
\label{eq:periodic:phi}
\eeqn

The solution~\eq{eq:phi:sum} for a single periodic HS instanton~\eq{eq:periodic:phi} corresponds to infinite amount of copies of the same radius $\lambda_k = \lambda$ with the positions $x_k = ({\bs x}_0, \tau_0 + \beta k)$. One gets for the HS scalar potential:
\beqn
\phi^{\mathrm{HS}}_{\beta} = 1 + \sum_{k \in {\mathbb Z}} \frac{\lambda^2}{({\bs x} - {\bs x}_0)^2 + (\tau - \tau_0 - k \beta)^2}, \qquad
\label{eq:phi:sum:T}
\eeqn
or, explicitly:
\beqn
\phi^{\mathrm{HS}}_\beta = 1 & + & \frac{\pi \lambda^2}{\beta} H({\bs x} - {\bs x}_0, \tau - \tau_0, \beta)\,, \qquad 
 \label{eq:HS:solution}
\eeqn
where we introduced the Harrington-Shepard function:
\beqn
H({\bs x}, \tau, \beta) & = & \frac{1}{|{\bs x}| } \frac{\sinh (2 \pi |{\bs x}|/\beta)}{\cosh (2 \pi |{\bs x}|/\beta) - \cos(2 \pi \tau/\beta)}.
\quad\
\label{eq:HS:function}
\eeqn

In the low-temperature limit, $\beta \to \infty$, the HS solution~\eq{eq:HS:solution} reduces to a zero-temperature single-instanton configuration with the topological charge $Q = \pm 1$:
\beqn
\lim_{\beta \to \infty} \phi^{\mathrm{HS}}_\beta = \phi^{\mathrm{inst}} \equiv 1 + \frac{\lambda^2}{({\bs x} - {\bs x}_0)^2 + (\tau - \tau_0)^0}\,,
\label{eq:instanton:limit}
\eeqn
In the opposite limit of high temperatures, 
\beqn
\lim_{\beta \to \infty} \phi^{\mathrm{HS}}_\beta = \phi^{\mathrm{mon}}_{\mathrm{BPS}} \equiv 1 + \frac{\pi \lambda^2}{\beta} 
\frac{1}{|{\bs x} - {\bs x}_0| }\,,
\label{eq:high:T:HS}
\eeqn
one finds the asymptotics of a dyon~\cite{Rossi:1978qe}: the caloron solution loses its dependence on the time coordinate and becomes a static, three-dimensional dyon-like configuration in the Bogomolny-Prasad-Sommerfield (BPS) limit~\cite{Bogomolny:1975de,Prasad:1975kr}. Despite its explicit self-duality, the BPS solution and its asymptotic are often referred to as a BPS monopole. One can also obtain the BPS expression~\eq{eq:high:T:HS} directly from Eq.~\eq{eq:HS:solution} replacing over the individual instantons by the integral, $\sum_n \to \int d n$, and performing the integration explicitly.

Monopole degrees of freedom provide the basis for one of the most popular mechanisms of color confinement which suggest that the confining force between a quark and an anti-quark appears due to a condensate of monopole-like gluonic configurations. At high temperature, the condensate evaporates and the color confinement disappears in agreement with the existing phenomenology~~\cite{Mandelstam:1974pi,tHooft:1981bkw}. This scenario, however, cannot work with the static BPS monopoles inherent to the high-temperature limit since any ensemble of the static monopoles contribute trivially to the confining order parameter, the Polyakov loop $P_{\bs x} = \exp\{\int_0^\beta d \tau\, A_4({\bs x}, \tau)\}$ (and at high temperature, the confining property is lost anyway). The simplest instanton-gas picture of gluon vacuum cannot explain the color confinement as well~\cite{Diakonov:1989un}.

At finite temperature, the BPS monopoles reveal themselves explicitly~\cite{Kraan:1998sn} as constituents of the so-called Kraan-van~Baal-Lee-Lu (KvBLL) calorons~\cite{Kraan:1998kp,Kraan:1998pm,Lee:1998vu,Lee:1998bb}. The KvBLL caloron is a generalization of the periodic HS instanton~\cite{Harrington:1976dj,Harrington:1978ve} extended to configurations that possess a nontrivial holonomy at spatial infinity, $P_{\infty} \equiv \lim_{x \to \infty} P_{\bs x} \neq \bbbone$. A gas of the KvBLL calorons has been shown to produce the confinement property~\cite{Gerhold:2006sk} similarly to confining models based on dyon configurations \cite{Diakonov:2007nv,Liu:2015ufa}

Given the promising applications of instantons, calorons, and their holonomically twisted counterparts to chiral symmetry breaking and confinement of QCD, in this article, we construct the finite-temperature instanton (caloron) solutions in the rotating Yang-Mills vacuum regarding their possible relevance to vortical quark-gluon plasma. We work with the case of trivial holonomy, $P_{\infty} \equiv \bbbone$, and consider Yang-Mills theory with $SU(2)$ gauge group. The generalization to $SU(N_c)$ is straightforward.

Consider a system that rotates with the angular velocity $\Omega$ about the axis $z$. In cylindrical coordinates of the inertial laboratory frame in Minkowski spacetime, $(\rho,\varphi,z,t)$, a uniform rotation of an object is encoded in the uniform growth (modulo $2\pi$) of the polar coordinate, ${\tilde \varphi} = {[\varphi - \Omega t]}_{2\pi}$ where the tilded coordinate refers to the system co-rotating with the object. The causality principle restricts the maximal extent of the object in the radial coordinate, $|\Omega \rho| < 1$ to guarantee the positivity of the purely time component of the metric $g_{tt}(\rho) = 1 - \Omega^2 \rho^2$ (we remind that $c=1$ in our units). 

After a Wick rotation to the Euclidean spacetime, the time coordinate becomes imaginary. The rotation becomes imaginary as well: the angular frequency $\Omega$ becomes a purely imaginary parameter, $\Omega_I = - i \Omega$~\cite{Yamamoto:2013zwa,Braguta:2020biu,Chernodub:2020qah,Braguta:2021jgn,Chen:2022smf}. The imaginary nature of rotation makes it possible to avoid the sign problem and opens a way for first-principle lattice simulations of rotating systems~\cite{Yamamoto:2013zwa,Braguta:2020biu,Braguta:2021jgn}. Under the Wick transformation, the real rotation in the real-time maps to the imaginary rotation with the imaginary angular frequency $\Omega_I$ in the imaginary time $\tau$ such that $\Omega t \to \Omega_I \tau$. In terms of the polar coordinate, the rotation in Euclidean space becomes ${\tilde \varphi} = {[\varphi - \Omega_I \tau]}_{2\pi}$, implying that the periodic constraint for the gluon field~\eq{eq:periodic:A} should be replaced, in a rotating medium, by the rotationally twisted periodicity:
\beqn
A_\mu(\rho,\varphi,z,\tau) = A_\mu(\rho,\varphi - \Omega_I \beta,z,\tau + \beta)\,.
\label{eq:twisted:A}
\eeqn

The uniform imaginary rotation with arbitrarily large $\Omega_I$ does not violate causality because there is no notion of causality in the Euclidean spacetime. In other words, the light cone does not exist in the Euclidean space (see also a discussion in Ref.~\cite{Chen:2022smf}). Moreover, at finite temperature, the theory is periodic in $\Omega_I$ since the spin-statistic relations for bosonic and fermionic theories imply the equivalence of the thermal state with respect to the following transformations~\cite{Chernodub:2020qah}: 
\beqn
\Omega_I & \to &\Omega_I + 2 \pi \beta^{-1} k, \qquad \mbox{(bosons)}, \label{eq:periodicity:bosons} \\
\Omega_I & \to &\Omega_I + 4 \pi \beta^{-1} k, \qquad \mbox{(fermions)}, \label{eq:periodicity:fermions}
\eeqn
where $k \in {\mathbb Z}$. The first relation for bosonic fields is seen from Eq.~\eq{eq:twisted:A}. For fermionic fields possessing anti-periodic boundary conditions, the system returns to its state after a double rotation. Interestingly, at $\Omega_I = 2 \pi \beta^{-1}$, the fermions become ghost particles as they are converted to spinors obeying bosonic statistics~\cite{Chernodub:2020qah}. 

How the imaginary rotation affects the periodic instanton (caloron)? The scalar potential $\phi$ is now determined by the rotationally twisted condition~\eq{eq:twisted:A} which replaces the finite-temperature periodic constraint~\eq{eq:periodic:phi}:
\beqn
\phi(\rho,\varphi,z,\tau) = \phi(\rho,\varphi - \Omega_I \beta,z,\tau + \beta)\,.
\label{eq:twisted:phi}
\eeqn

In order to satisfy Eq.~\eq{eq:twisted:phi} we supplement the sum~\eq{eq:phi:sum:T} with the rotational twists. A single caloron under the imaginary rotation is thus described by the potential:
\beqn
\phi_{\beta,\Omega_I} = 1 + \lambda^2 \sum_{k \in {\mathbb Z}} & & \bigl[\rho^2 + \rho_0^2 
- 2 \rho \rho_0 \cos(\varphi -\varphi_0 + \Omega_I \beta k)
\nonumber \\
& & + (z - z_0)^2 + (\tau - \tau_0 - \beta k)^2\bigr]^{-1}, \qquad
\label{eq:phi:twist:T}
\eeqn
where we use the cylindrical coordinate system and consider the system rotating about the axis $z$ centered at the origin $\rho = 0$. The original 4-position of the ``seed'' instanton is $x_0^\mu = (\rho_0,\varphi_0,z_0,\tau_0)$. Since we will be considering mainly the solutions seeded by a single instanton, we put, for simplicity, the seed instanton at the non-negative part of the $x$ axis, $\varphi_0 = 0$. A generalization to the many-instanton case is straightforward and follows identically the general prescription of Eq.~\eq{eq:phi:sum}.

In the low-temperature limit, $\beta \to \infty$, the rotated instanton~\eq{eq:phi:twist:T} reduces to the one-instanton solution~\eq{eq:instanton:limit} regardless of the value of $\Omega_I$. 

If the instanton is centered at the rotation axis, $\rho_0 = 0$, then this spherically symmetric system is insensitive to the rotation, and the solution~\eq{eq:phi:twist:T} reduces to the caloron solution given by the finite-temperature periodic instanton~\eq{eq:HS:solution}. Below we consider several nontrivial cases with $\rho_0 \neq 0$.

The explicit form of the potential~\eq{eq:phi:twist:T} can also be given for a set of imaginary angular frequencies $\Omega_I = 2 \pi \beta^{-1} \varkappa$ where $\varkappa$ is a rational number. Obviously, at $\Omega_I = 2\pi/\beta$ ($\varkappa = 1$), the system is insensitive to rotation due to the bosonic symmetry~\eq{eq:periodicity:bosons}. The first non-trivial case is given by $\Omega_I  = \pi/\beta$ ($\varkappa = 1/2$), which gives us $\cos(\varphi + \Omega_I \beta k) = (-1)^k \cos\varphi$ in the denominator of the sum~\eq{eq:phi:twist:T}. The sum in Eq.~\eq{eq:phi:twist:T} thus decouples into two sums over even and odd $k$ with the result:
\beqn
\phi_{\beta,\Omega_I} {\biggl|}_{\Omega_I = \frac{\pi}{\beta}} = 1 & + & \frac{\pi \lambda^2}{\beta} \bigl[ H({\bs x} - {\bs x}_0, \tau - \tau_0, 2 \beta) \nonumber \\
& & + H({\bs x} + {\bs x}_0, \tau - \tau_0 + \beta, 2 \beta) \bigr]\,, \qquad 
\label{eq:twiston:N2}
\eeqn
where $H$ is the HS function~\eq{eq:HS:function}. The sense of Eq.~\eq{eq:twiston:N2} is clear: the $\pi$-rotating (flipping) instanton chain consists of two sub-chains that can be summed over independently. This example provides us with the idea of how to build the rotating instanton solution for any rational (in units of $2 \pi/\beta \equiv 2 \pi T$) angular frequency:
\beqn
\Omega_I = \frac{2 \pi}{\beta} \frac{m}{n}\,, 
\quad 
m = 0, \dots n - 1,
\quad 
n = 1,2, \dots\,, \quad
\label{eq:rational:Omega:I}
\eeqn
where the ratio $m/n$ represents an irreducible fraction.

The general solution for the angular frequency~\eq{eq:rational:Omega:I} is:
\beqn
\phi_{\beta,\Omega_I} {\biggl|}_{\Omega_I  = \frac{2\pi}{\beta} \frac{m}{n}} \!\! & = &  1 + \frac{\pi \lambda^2}{\beta} 
\label{eq:twiston:n} \\
& & 
\times \sum_{l=0}^{n-1} H\bigl({\bs x} - {\bs {\bar x}}_n^{(lm)}, \tau - \tau_0 + l \beta, n \beta\bigr), \quad \nonumber 
\eeqn
where we introduced the set of images ${\bs {\bar x}}_n^{(k)}$ of the position of the seed instanton, ${\bs x}_0 = (x_0 \equiv \rho_0,0,z_0)$, rotated about the $z$ axis by the angle $2\pi k/n$: 
\beqn
{\bs {\bar x}}_n^{(k)} = \left(x_0 \cos \frac{2 \pi k}{n}, \ x_0 \sin \frac{2 \pi k}{n}, \ z_0 \right)^T\,.
\label{eq:xk}
\eeqn
In the simplest nontrivial case, ${\bs {\bar x}}_2^{(1)} = - {\bs {\bar x}}_2^{(0)} \equiv - {\bs x}_0$ enters the $n=2$ solution~\eq{eq:twiston:N2} as the mirror image of the position ${\bs x}_0$ of the seed instanton.

\begin{figure*}[!thb]
\centering
\includegraphics[width=0.95\linewidth]{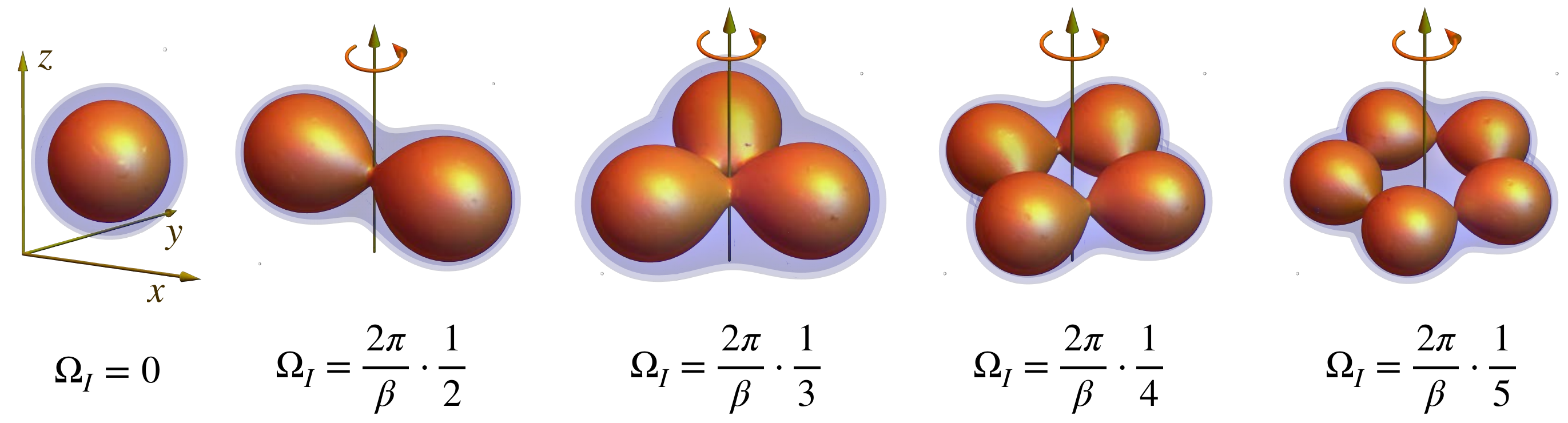}
\caption{Equipotential lines of the scalar potential~\eq{eq:twiston:n} of the rotating caloron solution at fixed $\tau = \beta/2$ and various imaginary angular frequencies~$\Omega_I$.}
\label{fig:mballs}
\end{figure*}

The geometric localization of the topological charge~\eq{eq:Q} and action~\eq{eq:S} densities in the classical self-dual configurations closely follows the behavior of the scalar function~$\phi$. In order to get an insight into the effect of the rotation on the localization properties of the solutions, we show in Fig.~\ref{fig:mballs} the scalar potential $\phi$ in the spatial volume for various rational frequencies $\Omega_I$. 

The rotation produces several fascinating effects:
\begin{enumerate}
\item First of all, the imaginary rotation with the rational frequency generates exactly $n$ identical lumps over the spatial ring, where $n$ is the denominator of the irreducible fraction $m/n$ of the frequency~\eq{eq:rational:Omega:I}. Since the lumps represent identical structures, their charge is fractionalized. Each lump carries the topological charge of $1/n$ of the original solution.

\item The number of the lumps does not depend on the numerator $m$ that determines the rational frequency~\eq{eq:rational:Omega:I}. This property implies that the rotation with the frequencies, for example, $\Omega_I = \frac{1}{5} \cdot 2 \pi T$ and $\Omega_I = \frac{2}{5}\cdot 2 \pi T$ lead to the identical solutions. 

\item As one can see from the analytical formula~\eq{eq:twiston:n}, the rotation with the rational $m/n$ frequency effectively cools down the caloron configuration by reducing its temperature from $T$ down to $T/n$. 
\end{enumerate}

Thus, the imaginary rotation with the rational (in units of $2 \pi T$) angular frequency leads to the fractionalization and delocalization of the topological charge of the finite-temperature caloron and the temperature drop in the solution. The latter property might imply that the imaginary rotation effectively decreases the temperature in the topological sector of the gluonic plasma. 

We suggest that a rotating caloron at an irrational frequency (for example, at $\Omega_I = \sqrt{2} \pi T$ or $\Omega_I = 0.1 T$) reduces, regardless of the temperature $T$, to a zero-temperature configuration with an infinite number of instantons placed along a circular ring of a finite radius. The radius is equal to the distance from the seed instanton's center to the rotation axis. This ``instanton-ring'' solution, known to produce a monopole along the ring~\cite{Bruckmann:2003bd}, can provide us with another intriguing link between rotation and the problem of color confinement.

The strong sensitivity of the basic properties of the rotating calorons, such that the effective temperature and the localization of topological charge, calls for a careful study of the analytical continuation of observables from the imaginary to real angular frequencies, $\Omega_I \to \Omega$, at least in the topological sector of Yang-Mills theory. For example, the angular frequencies $\Omega_I = 2 \pi T/10$ and $\Omega_I = 2 \pi T \cdot 99/1000$, that differ only by $1\%$, produce substantially different caloron configurations that feature, respectively, 10 and 1000 constituents each of which carries 1/10 and 1/1000 fraction of the unit topological charge. 

Finally, let us consider the high-temperature limit of the rotating instanton solution. In the non-rotating case, the HS periodic instanton~\eq{eq:HS:solution} reaches its BPS monopole limit~\eq{eq:high:T:HS} when the inverse temperature $\beta$ becomes smaller than the typical length scale $\lambda$ of the solution, $\beta \ll \lambda$. For solution~\eq{eq:HS:solution}, this scale corresponds to a smaller length out of the two length parameters, $\lambda \sim |x-{\bs x}_0| \sim |\tau - \tau_0|$, which in dynamical configurations, are dictated by the scale factor $\lambda$ in Eq.~\eq{eq:phi:sum:T}.

The rotating instanton~\eq{eq:twiston:n} is characterized by three parameters: its size $\lambda$, the inverse temperature scale $\beta$, and the imaginary angular frequency $\Omega_I$. While the hierarchy $\beta \ll \lambda$ is rigidly fixed by the requirement of the high-$T$ limit, the properties of the high-$T$ solution may vary significantly depending on the value of the frequency~$\Omega_I$. 

In the case of slow rotation, the frequency is small compared to the length scales of the solution, $\Omega_I \lambda \ll 1$. Assuming rationality of the angular velocity~\eq{eq:rational:Omega:I}, the high-$T$ limit is achieved in the limit $m \beta \ll \lambda$, where $m$ is defined as the denominator in the irreducible fraction in Eq.~\eq{eq:rational:Omega:I}. The scalar potential of the slowly rotating high-temperature instanton is then obtained by replacing the HS functions in Eq.~\eq{eq:twiston:n} with their BPS limits~\eq{eq:high:T:HS}:
\beqn
\phi^{\mathrm{high}T}_{\mathrm{slow}} {\biggl|}_{\Omega_I  = \frac{2\pi}{\beta} \frac{m}{n}} \!\!\!
=  1 + \frac{\pi \lambda^2}{\beta} \frac{1}{|{\bs x} - {\bs {\bar x}}_n^{(m)}| }\,,
\label{eq:twiston:n:slow} 
\eeqn
where the monopole positions ${\bs {\bar x}}_n^{(k)}$ is given in Eq.~\eq{eq:xk}. Thus, the high-temperature limit of the slowly rotating solution~\eq{eq:twiston:n} is a simple sum of the static BPS monopoles. 

On the contrary, the limit of the rapidly rotating instanton, $\Omega_I \lambda \gg 1$, is a nontrivial example of an analytically solvable high-tem\-pe\-ra\-tu\-re ($\beta \ll \lambda$) solution provided\footnote{These three hierarchical relations are consistent with each other. For example, one can take $\beta = 0.1\,{\mathrm{fm}}$, $\lambda = 1\,{\mathrm{fm}}$, $\Omega_I = 10\,{\mathrm{fm}}^{-1}$.} $\Omega_I \beta \ll 2 \pi$. In this limit, the term $\Omega_I \beta k$ under the cosine function in the sum over~\eq{eq:phi:twist:T} represents a ``fast'' variable that oscillates more rapidly compared to a ``slow'' evolution of the rest of the expression. Performing first the integration averaging over a full period of the fast variable and then taking the integral over the slow variable, we get for the scalar potential of the rapidly rotating caloron the following analytical expression:
\beqn
\phi_{\circ}(\rho,z) =  1 + 
\frac{2\lambda^2 \, K\bigl(-\frac{4 \rho \rho_0}{(\rho - \rho_0)^2 
+ (z - z_0)^2}\bigr)}{\beta \sqrt{(\rho - \rho_0)^2 + (z - z_0)^2}}\,,
\label{eq:circulon} 
\eeqn
where $K(x)$ is complete elliptic integral of the first kind\footnote{In other words, $K(x) = \int_0^{\frac{\pi}{2}} \frac{d \theta}{\sqrt{1 - x^2 \sin^2 \theta}}$.}. We call the high-temperature rapidly-rotating caloron solution~\eq{eq:circulon} ``circulon''.

\begin{figure}[thb]
\centering
\includegraphics[width=0.95\linewidth]{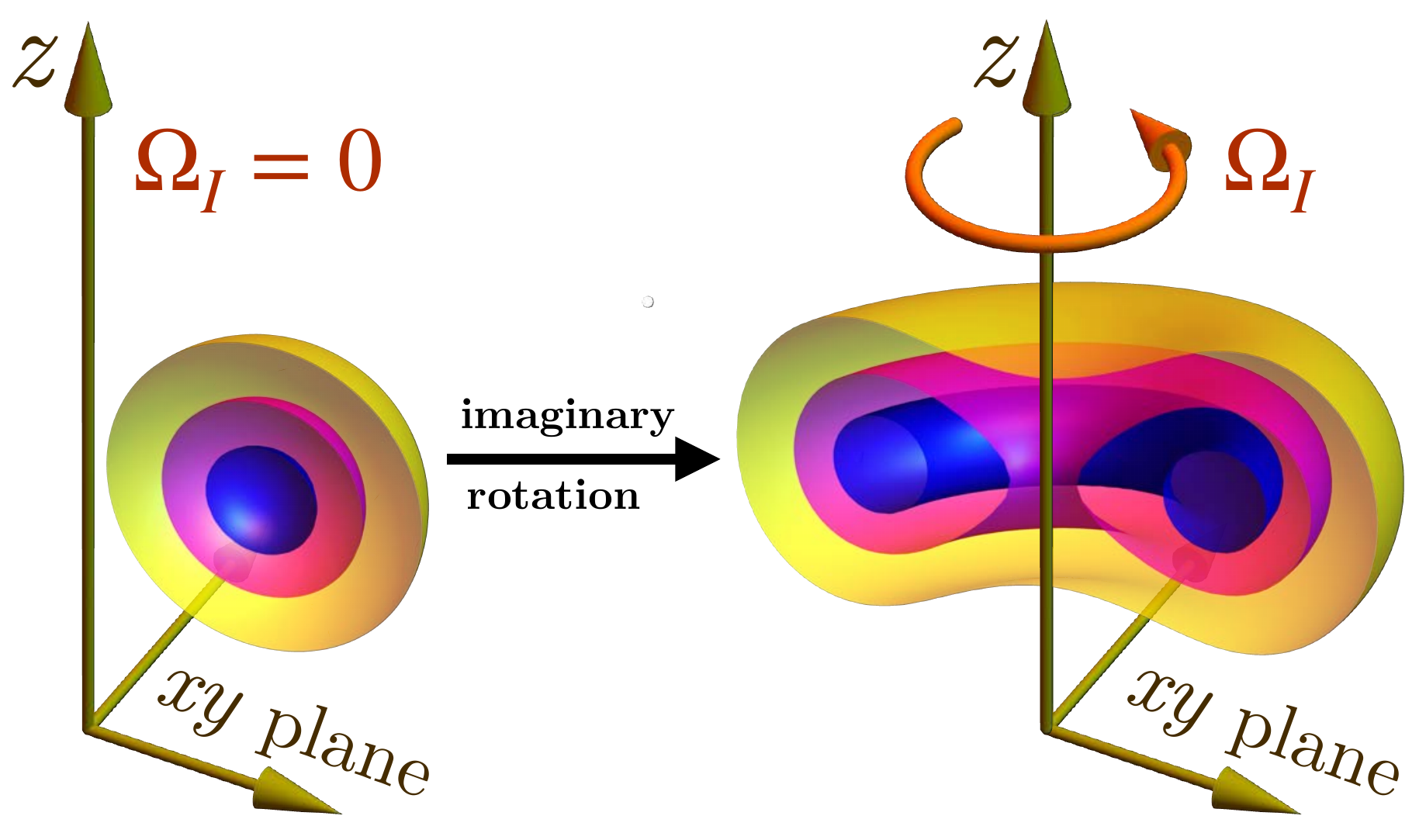}
\caption{Equipotential lines for the topological charge (action density) of (left) the HS caloron solution~\eq{eq:HS:solution}, \eq{eq:HS:function} and (right) the circulon solution~\eq{eq:circulon}. Fast imaginary rotation smears the caloron over the ring and produces the circulon.}
\label{fig:smearing}
\end{figure}

Remarkable, the circulon solution~\eq{eq:circulon} does not depend on the angular frequency $\Omega_I$ implying that Eq.~\eq{eq:circulon} is an ultimate configuration in the limit of a high-frequency rotation. Moreover, the high-$T$ high-$\Omega_I$ solution~\eq{eq:circulon} possesses the axial symmetry implying that the original caloron gets smeared uniformly by the rotation. The core of the circulon forms a ring at $(\rho,z) = (\rho_0,z_0)$, Fig.~\ref{fig:smearing}.

At large distances $r_\delta \equiv \sqrt{(\rho - \rho_0)^2 + (z - z_0)^2}$ from the ring $(\rho,z) = (\rho_0,z_0)$, the circulon potential~\eq{eq:circulon} acquires the asymptotic form of the BPS monopole~\eq{eq:high:T:HS}:
\beqn
\phi_\circ = 1 + \frac{\pi \lambda^2}{\beta} \frac{1}{r_\delta} + \dots, 
\qquad 
r_\delta \gg \rho_0\,,
\eeqn
where the ellipsis denotes the subleading $O(r^{-3}_\delta)$ terms.

The circulon potential~\eq{eq:circulon} has a logarithmic divergence close to its ring-shaped core:
\beqn
\phi_{\circ} = 1 - \frac{\lambda^2}{\beta \rho_0} \log\frac{r_\delta}{8 \rho_0} + \dots,
\qquad 
r_\delta \ll \rho_0\,,
\label{eq:short:distance}
\eeqn
where the ellipsis corresponds to $O(r^2_\delta)$ corrections. Therefore, the rapid rotation softens the point-like divergent pole of the static BPS solution~\eq{eq:high:T:HS} down to the logarithmic singularity~\eq{eq:short:distance}.

\begin{figure}[!thb]
\centering
\includegraphics[width=0.5\linewidth]{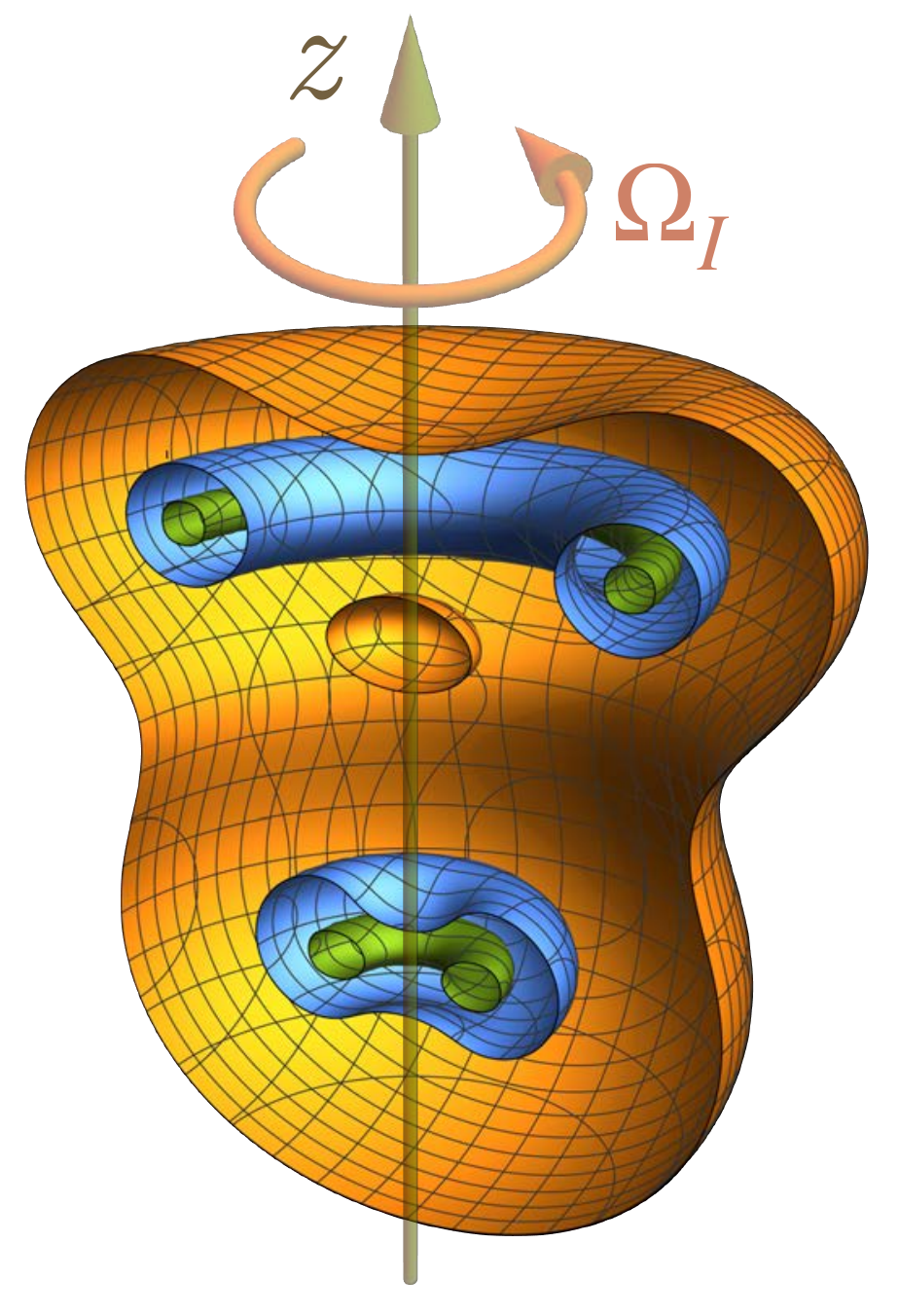}
\caption{Equipotential lines of the topological charge (action) density of a two-circulon solution with the cores at $(\rho_0,z_0)=(\lambda,0),\ (3\lambda,6\lambda)$ and inverse temperature $\beta = 0.1 \lambda$.}
\label{fig:happy}
\end{figure}

A generalization of our construction to many-circulon solutions is straightforward. An example of a serious two-circulon configuration is shown in Fig.~\ref{fig:happy}.

In summarizing, we constructed the finite-temperature instanton (caloron) solution in Euclidean Yang-Mills theory subjected to imaginary rotation. The solution with a rational-valued angular frequency $\Omega_I$ can be represented as an ensemble of the Harrington-Shepard calorons arranged equidistantly along a ring~\eq{eq:twiston:n}. We demonstrated that the imaginary rotation cools down the caloron solution, reducing its temperature and leading, at the same time, to the delocalization and subsequent fractionalization of its topological charge. In the high-temperature limit, a slowly rotating caloron reduces to an ensemble of static BPS monopoles. A fast imaginary rotation (which still belongs to a physically reasonable domain with $\Omega_I \ll 2 \pi T$) produces the ``circulon'' solution with a toroidal core. The circulon potential can explicitly be expressed via the elliptic integral~\eq{eq:circulon}. 


%

\end{document}